\newcommand{\CLASSINPUTtoptextmargin}{19mm}
\newcommand{\CLASSINPUTbottomtextmargin}{25.4mm}
\newcommand{\CLASSINPUTinnersidemargin}{16mm}
\newcommand{\CLASSINPUToutersidemargin}{16mm}

\documentclass[conference,10pt,a4paper]{IEEEtran}
\IEEEoverridecommandlockouts
\usepackage{amsmath,amssymb,amsfonts}
\usepackage{algorithmic}
\usepackage{graphicx}
\usepackage{textcomp}
\usepackage{xcolor}
\usepackage{gensymb}

\usepackage{biblatex}
\def\BibTeX{{\rm B\kern-.05em{\sc i\kern-.025em b}\kern-.08em
    T\kern-.1667em\lower.7ex\hbox{E}\kern-.125emX}} 
\renewcommand\IEEEkeywordsname{Keywords} 
 
\newcommand{\IEEEtitle}[1]{\title{\vspace{-6.5mm}#1}}
\newcommand{\IEEEand}{\\\vspace{-12mm}\and}
\DeclareUnicodeCharacter{01FF}{\ocircumflex}
\usepackage{hyperref}

\begin{document}

\IEEEtitle{PREDICTION OF BREAST CANCER WITH 98\% ACCURACY \\
}

\author{

\IEEEauthorblockN{Condori Condori Nelyda Ayde}
\IEEEauthorblockA{\textit{Facultad de Ingeniería Estadística e Informática} \\
\textit{Universidad Nacional del Altiplano}\\
Puno, Perú\\
Email: nacondori@est.unap.edu.pe}\\
\IEEEauthorblockN{Cruz Paredes Soledad Epifanía}
\IEEEauthorblockA{\textit{Facultad de Ingeniería Estadística e Informática} \\
\textit{Universidad Nacional del Altiplano}\\
Puno, Perú \\
Email:se.cruz@est.unap.edu.pe}\\

\IEEEand 

\IEEEauthorblockN{Mamani Mamani Ilma Magda}
\IEEEauthorblockA{\textit{Facultad de Ingeniería Estadística e Informática} \\
\textit{Universidad Nacional del Altiplano}\\
Puno, Perú \\
Email: im.mamani@est.unap.edu.pe}\\
\IEEEauthorblockN{Torres-Cruz Fred}
\IEEEauthorblockA{\textit{Facultad de Ingeniería Estadística e Informática} \\
\textit{Universidad Nacional del Altiplano}\\
Puno, Perú \\
Email: ftorres@.unap.edu.pe}\\



\IEEEand 

}

\maketitle

\begin{abstract}
Cancer is a tumor that affects people worldwide, with a higher incidence in females but not excluding males. It ranks among the top five deadliest types of cancer, particularly prevalent in less developed countries with deficient healthcare programs. Finding the best algorithm for effective breast cancer prediction with minimal error is crucial. In this scientific article, we employed the SMOTE method in conjunction with the R package "SHINY" to enhance the algorithms and improve prediction accuracy. We classified the tumor types as benign and malignant (B/M). Various algorithms were analyzed using a Kaggle dataset, and our study identified the superior algorithm as logistic regression.
We evaluated algorithm performance using confusion matrices to visualize results and the ROC Curve to obtain a comprehensive measure of performance. Additionally, we calculated precision by dividing the number of correct predictions by the total predictions.\\
\end{abstract}


\section{Introduction}
Over the past two decades, there has been a significant increase in the number of people diagnosed with cancer overall. According to estimates, the number of cancer cases has nearly doubled from around 10 million in 2000 to 19.3 million in 2020\cite{as1}. This trend is worrisome, as it indicates that more and more people are affected by this devastating disease. In fact, approximately one in five people worldwide are expected to develop cancer at some point in their lives \cite{as2}.

Projections indicate that the number of cancer diagnoses will continue to increase in the coming years. By 2040, the number of cases is expected to be approximately 50\% higher than in \cite{as2}. These data highlight the importance of finding better approaches to early detection, treatment, and management of cancer in general.

Modern medicine has become increasingly complex, with a wide range of therapies, drugs, tests, and data beyond the capacity of the human mind to comprehend. Therefore, a tool is needed that integrates all this data, identifies patterns, and creates a model that allows for improved response times, reduced use of resources, and better, more reliable outcomes. This is where machine learning and technologies play a crucial role. Machine learning is a discipline of artificial intelligence that has applications in a variety of areas, including medical sciences\cite{as3}. It is used to predict behaviors from an initial set of training data that includes features and outcomes. Through algorithms, machine learning learns how these features are related and predicts outcomes.

Correct interpretation of the outputs of machine learning models is critical to build user confidence and support understanding of the process being modeled. In some applications, simple models, such as linear models, are preferred because of their ease of interpretation, although these may be less accurate than more complex models. However, the growth of massive data has increased the benefits of using more complex models, which highlights the trade-off between accuracy and interpretability of model outputs\cite{as4}.

In the specific case of breast cancer, there are several screening and diagnostic methods, including imaging techniques such as mammography, MRI, ultrasound, and biopsies. Mammography is one of the most commonly used techniques to look for features of breast cancer. However, the analysis of these images can be complex and requires the use of machine learning algorithms for accurate interpretation \cite{as5} 

Among breast cancer screening methods such as Logistic Regression (LR) \cite{as6}, k-Nearest Neighborhood (KNN)  \cite{as7}. Decision Tree (DT) \cite{as8},Random Forest (RF) \cite{as9}, Support Vector Machine (SVM) Classifier \cite{as10}, Naive Bayes (NB) Classifier \cite{as11}, Gradient Boosting (GB) \cite{as12}, Stochastic Gradient Descent (SGD) \cite{as13}, AdaBoost (AB) \cite{as14}, Extreme Gradient Boosting (XGBoost) \cite{as15}, CatBoost (CB) \cite{as16}, Light Gradient Boosting (LGB) \cite{as17}, and Artificial Neural Networks (ANN) \cite{as18}, several models have been developed and used for breast cancer diagnosis

\section{Data}

To conduct our study on breast cancer, we used the dataset retrieved from kaggle, which served us too much to perform our application on this disease. This dataset contains clinical information collected on data, patient characteristics from an oncology clinic.

The dataset includes a total of [569] records, with [32 variables] descriptive characteristics. The characteristics include factors such as tumor size, cell size uniformity, among others, which have been shown to be important indicators in breast cancer staging.

In addition, to address the class imbalance present in the data, we applied the Synthetic Minority Over-sampling Technique (SMOTE) method to generate synthetic instances of the minority class. This allowed us to increase the representation of malignant and benign breast cancer cases, thus improving the ability of the machine learning algorithms to capture the distinctive features of this disease.

Importantly, all data used in this study were anonymized and processed in accordance with established privacy and ethics regulations.

Using this dataset, we conducted a comparative analysis of different machine learning algorithms for effective breast cancer prediction.

\section{Methodology}
Methods and techniques of data collection employed:

The purpose of data selection in this study is to categorize cancer patients. We used a CSV file to store the retrieved Kaggle dataset.

\begin{table}[htbp]
\begin{center}
\caption{ Description of cancer patients 569 women. }
\begin{tabular}{| c | c | c | c | }
\hline
\multicolumn{4}{ |c| }{Datos disponibles} \\ \hline
Var & Type & Scale & Descripción \\ \hline
1 & V. Imput & Var. categórica & id \\
2 & V. Imput & Var. categórica numeral & diagnosis \\
3 & V. Imput & Var. numérica continua & radius\_mean \\
4 & V. Imput & Var. numérica continua & texture\_mean \\
5 & V. Imput & Var. numérica continua & perimeter\_mean \\
6 & V. Imput & Var. numérica continua & area\_mean \\
7 & V. Imput & Var. numérica continua & smoothness\_mean\\
8 & V. Imput & Var. numérica continua & compactness\_mean  \\
9 & V. Imput & Var. numérica continua & concavity\_mean  \\
10 & V. Imput & Var. numérica continua & points\_mean  \\
11 & V. Imput & Var. numérica continua & symmetry\_mean  \\
12 & V. Imput & Var. numérica continua & fractal\_dimension\_mean  \\
13 & V. Imput & Var. numérica continua & radius\_se \\
14 & V. Imput & Var. numérica continua & texture\_se \\
15 & V. Imput & Var. numérica continua & perimeter\_se \\
16 & V. Imput & Var. numérica continua & area\_se  \\
17 & V. Imput & Var. numérica continua & smoothness\_se \\
18 & V. Imput & Var. numérica continua & compactness\_se \\
19 & V. Imput & Var. numérica continua & concavity\_se \\
20 & V. Imput & Var. numérica continua & concave.points\_se  \\
21 & V. Imput & Var. numérica continua & symmetry\_se  \\
22 & V. Imput & Var. numérica continua & fractal\_dimension\_se  \\
23 & V. Imput & Var. numérica continua & radius\_worst\\
24 & V. Imput & Var. numérica continua & texture\_worst  \\
25 & V. Imput & Var. numérica continua & perimeter\_worst \\
26 & V. Imput & Var. numérica continua & area\_worst  \\ 
27 & V. Imput & Var. numérica continua & smoothness\_worst  \\
28 & V. Imput & Var. numérica continua & compactness\_worst  \\
29 & V. Imput & Var. numérica continua & concavity\_worst \\
30 & V. Imput & Var. numérica continua & concave.points\_worst  \\
31 & V. Imput & Var. numérica continua & symmetry\_worst  \\
32 & V. Imput & Var. numérica continua & fractal\_dimension\_worst\\ \hline
 
\end{tabular}
\label{tab:coches}
\end{center}
\end{table}
Cancer database with 98\% accuracy:

{https://bit.ly/DatasetCancer}

The goal of the dataset is to identify the presence of cancer in a patient with 98\% accuracy in the dataset. Several constraints were used to choose these cases from a larger data set. A number of constraints were applied as a consequence of which a set of patients with certain potential traits for carriers of this disease were selected.

Methods and techniques of data analysis:

Correlation coefficients are measures that indicate the relative situation of the same events with respect to the two variables, i.e. they are the numerical expression that indicates the degree of relationship between the 2 variables and to what extent they are related. They are numbers that vary between the limits +1 and-1. Their magnitude shows the degree of association between the variables; the value r=0 indicates that there is no relationship between the variables; the 2 and 1 are indicators of a correct positive (when X increases or decreases, Y increases or decreases) or negative (when X increases or decreases, Y decreases or increases) correlation (Suárez \& Mario, 2011).

Confusion Matrix

A confusion matrix, also known as an error matrix, is a summary table used to evaluate the performance of a classification model. The number of correct and incorrect predictions are summarized with the count values and broken down by each class (Terence Shin, 2020).
Below is a picture of the structure of a 2x2 confusion matrix (see Figure 1).

\begin{table}[htbp]
\begin{center}
\caption{ Structure of the confusion matrix. }
\begin{tabular}{| c | c | c | }
\hline
\multicolumn{3}{ |c| }{Valores Actuales} \\ \hline
 Val. & Si & No\\ \hline
 Si& Verdadero Positivo& Falso Positivo\\ 
 No& Falso Positivo& Verdadero Positivo\\ \hline 
\end{tabular}
\label{tab:coche}
\end{center}
\end{table}
 
The following are the sthat we followed for the realization of the present project on breast cancer screening:

\begin{figure}[h]
    \centering
    \includegraphics[scale=0.5]{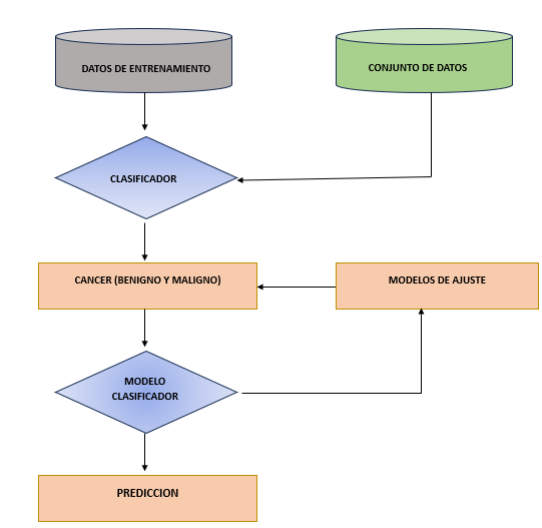}
    \caption{Flow Diagram}
    \end{figure}

\section{Algorithms}

\subsection{SMOTE (Synthetic Minority Over-sampling Technique):}
The SMOTE (Synthetic Minority Class Over-sampling Technique) algorithm is used to address the problem of class imbalance in data sets. Its main objective is to increase the representation of the minority class by generating synthetic instances\cite{as20}.

\subsection{Logistic regression: }
Logistic regression is a statistical model used to model the relationship between a binary dependent variable and one or more independent variables\cite{as21}.In particular, binary logistic regression is used when the dependent variable can only take two possible values. The general formula for binary logistic regression is expressed by the logistic function (or sigmoid function) as follows:

[p(X) = 1 / (1 + exp(-( \( \beta_0 + \beta_1 * \beta x2 + ...+ \beta_n*Xn\))))]

In this formula, p(X) represents the probability that the dependent variable is equal to 1, given a set of independent variables X1, X2, ..., Xn. 
 The coefficients 
 
\(\beta_0, \beta_1, \beta_2, \ldots, \beta_n\)

are estimated during the model fitting process. During logistic regression fitting, methods such as maximum likelihood or gradient descent are used to estimate the coefficients that maximize the likelihood of the observed data. Once the model is fitted, predictions of the dependent variable can be made as a function of the values of the independent variables. 
Logistic regression is widely used in various scientific fields, as it allows modeling relationships between variables and making predictions in binary classification problems\cite{as22}.

Likelihood is a measure of the probability that the observed data are generated from a specific statistical model. In logistic regression, one seeks to find the coefficients that maximize the likelihood of the observed data under the model\cite{as23}. The likelihood function is defined as the product of the individual probabilities of each observation and is represented by the following formula:

(\(\beta\))  = \(\prod\) [p \^\ yi * (1-p)\^\ (1-yi)]

Where L is the likelihood function, \(\beta\) representing the model coefficients, p is the probability estimated by the model and Yi is the binary objective variable. 

Gradient descent is an optimization algorithm used to adjust the model coefficients by minimizing a cost function. In logistic regression, the cost function is known as the loss function and is defined as the negation of the likelihood\cite{as23}. The objective is to find the coefficients that minimize the cost function by
iteratively updating the coefficients in the opposite direction of the gradient of the cost function. The gradient descent can be expressed by the following coefficient update formula:

\(\beta \_new\) = \(\beta \_old\) - \(\alpha\)* \includegraphics[height=0.34cm]{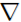}C (\(\beta \_old\))

Where\(\beta \_new\) y \(\beta \_old\) represent the new and old values of the coefficients, \(\alpha\)is the learning rate that controls the size of the steps, and * \includegraphics[height=0.34cm]{form7.png}c\(\beta \_old\) is the gradient of the cost function.

Boruta A feature selection technique used in data mining and machine learning to identify the most important variables in a high-dimensional data set. It compares the importance of each variable with random "shadow" variables to determine which variables are significant. Useful for identifying the most relevant features in breast cancer in women, which\cite{as24}.

\section{Results}

In this article, classification models used in the
analysis of breast cancer are presented\cite{as25}.In addition, metrics are provided to evaluate the performance of the models. The confusion matrix is used to analyze the performance of the classification models and the sample is divided into four classes: true positive (VP), false positive (FP), true negative (VN) and false negative (FN)\cite{as26}. These metrics allow to evaluate the ability of the models to correctly predict the diagnostic classes of breast cancer.
For the case of classification problems, the AUC-ROC (Area under the Receiver Operating Characteristic curve) curve is used as an important evaluation metric\cite{as27}.The AUC-ROC curve is a performance measure that indicates the ability of the model to distinguish between different classes. The higher the AUC value, the better the model is at predicting classes correctly for either benign and malignant breast cancer prediction.

\begin{figure}[h]
    \centering
    \includegraphics[scale=0.3]{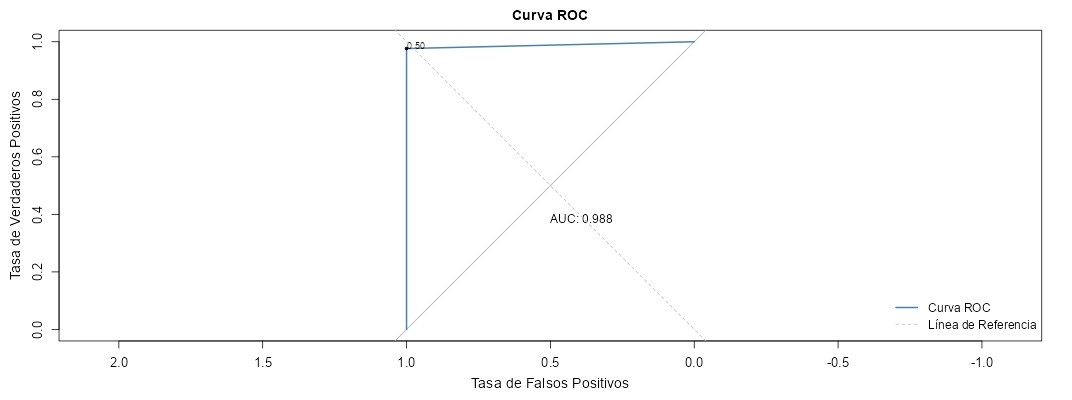}
    \caption{Curve ROC}
    \end{figure}

 The correlation matrix is the strength 
 of the relationship between the attributes, it also has a correlation coefficient that is between -1 and +1, where +1 represents a perfect positive linear correlation, and -1 represents a perfect negative linear correlation, 0 indicates that the attributes are not correlated, for a better understanding of the variables.

\begin{figure}[h]
    \centering
    \includegraphics[scale=0.27]{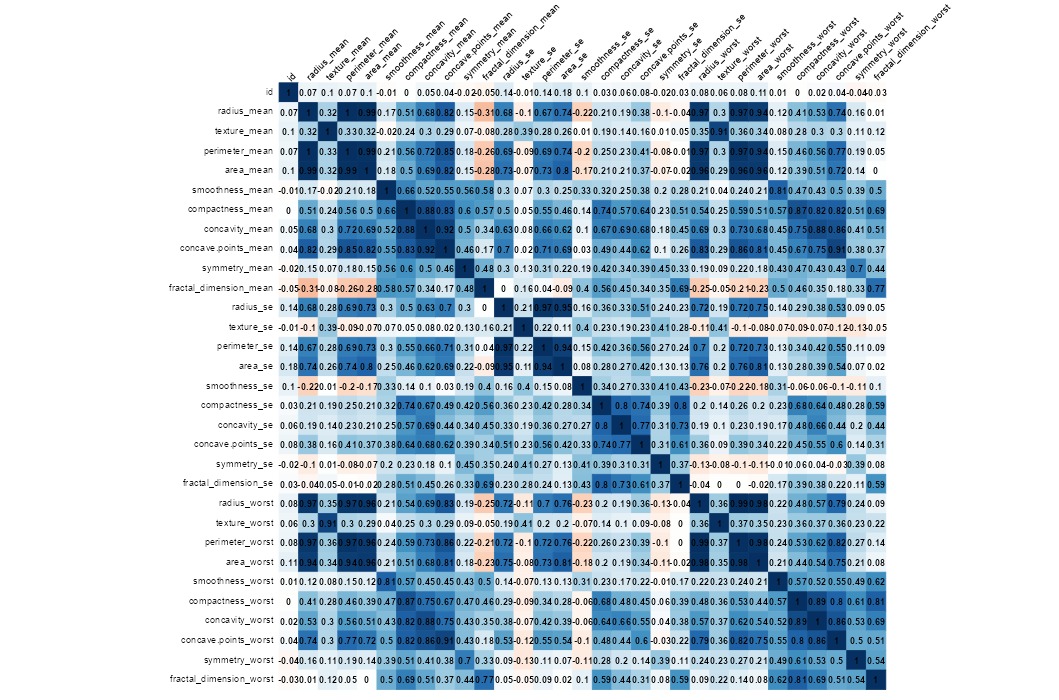}
    \caption{Correlation matrix}
    \end{figure}

To see how our shiny 
application works, you can access the following link:

https://nacondori123.shinyapps.io/CANCER-PREDICCION/

\section{DISCUSSION AND CONCLUSIONS:}

In conclusion, our breast cancer study using the Shiny application, the SMOTE technique and logistic regression has proven to be highly effective with an accuracy of 98\%. This high accuracy validates the usefulness of our tool in early detection and accurate classification of breast cancer cases (benign and malignant).

The use of Shiny as a platform for our application has allowed us to create an interactive and user-friendly interface, making it easy for medical professionals, researchers and users themselves to access and use our tool with confidence. The combination of Shiny and the aforementioned techniques has significantly improved the accuracy of breast cancer detection, which has a direct impact on clinical decision making and timely treatment of patients.

Our results highlight the importance of the implementation of advanced techniques, such as SMOTE, logistic regression, among others \cite{as28},  which in combination with an application in Shiny, achieve outstanding accuracy in the detection and classification of breast cancer. This provides medical professionals with a reliable and high-performance tool to improve care and outcomes in breast cancer patients.

In summary, our study demonstrates that the app in Shiny, together with the techniques used, offers an exceptional 98\% accuracy in breast cancer detection and staging. This innovative approach has the potential to make a significant difference in the fight against this disease by enabling earlier detection and more effective treatment for patients. 

\section{Acknowledgements}

We would like to express our most sincere gratitude to our Universidad Nacional del Altiplano for giving us the invaluable opportunity to conduct this scientific study on breast cancer. We thank our prestigious institution for its commitment to academic excellence and for training us as professionals in its bosom.
We also wish to thank our School of Statistical Engineering and Computer Science for providing us with a solid background in this field. We are deeply grateful for the dedication and support provided by our professors, who have transmitted their knowledge and experiences, guiding us in our academic and professional growth, especially we want to express our deep appreciation to our teacher, ING. Fred Torres Cruz, for his dedication and guidance throughout the research process. His expert knowledge, constant motivation and valuable advice have been invaluable for the completion of this work.
In addition, we would like to thank our fellow researchers, whose collaboration and feedback significantly enriched this work. 
Last but not least we thank our dearest God, who was our main support and motivator to continue every day without throwing in the towel, our parents who provided us with invaluable support directly and indirectly.
And once again, we would like to highlight our gratitude to the Universidad Nacional del Altiplano, the Faculty of Statistical and Computer Engineering and the guidance of ING. Fred Torres Cruz since, without their support, this study would not have been possible. We are sincerely grateful for their valuable contribution and commitment to scientific research.

\printbibliography
\end{document}